\documentclass[preprint]{elsarticle}
 
\usepackage{amssymb}
\usepackage{graphicx}
\usepackage{dcolumn}
\usepackage{bm}
\usepackage{setspace}
\usepackage{here}
\usepackage{slashbox}
  
\newcommand{\met}{\not\!\!E_T}

\def\InvMtt{\sqrt{\hat{s}}}
\def\qqtt{q\bar{q}\rightarrow t\bar{t}}
\def\InvMttSQp{{\hat{s}_p}}

\journal{Physics Letters B}
 
\begin{document}

\begin{frontmatter}
 
\title{Search for New Color-Octet Vector Particle Decaying to $t\bar{t}$ in $p\bar{p}$ Collisions at $\sqrt{s}=1.96$ TeV}

\address[AcademiaSinica]{Institute of Physics, Academia Sinica, Taipei, Taiwan 11529, Republic of China}
\address[Argonne]{Argonne National Laboratory, Argonne, Illinois 60439, United States}
\address[Athens]{University of Athens, 157 71 Athens, Greece}
\address[Barcelona]{Institut de Fisica d'Altes Energies, Universitat Autonoma de Barcelona, E-08193, Bellaterra (Barcelona), Spain}
\address[Baylor]{Baylor University, Waco, Texas  76798, United States}
\address[Bologna]{Istituto Nazionale di Fisica Nucleare Bologna, University of Bologna, I-40127 Bologna, Italy}
\address[Brandeis]{Brandeis University, Waltham, Massachusetts 02254, United States}
\address[UCDavis]{University of California, Davis, Davis, California  95616, United States}
\address[UCLA]{University of California, Los Angeles, Los Angeles, California  90024, United States}
\address[UCSanDiego]{University of California, San Diego, La Jolla, California  92093, United States}
\address[UCSantaBarbara]{University of California, Santa Barbara, Santa Barbara, California 93106, United States}
\address[Cantabria]{Instituto de Fisica de Cantabria, CSIC-University of Cantabria, 39005 Santander, Spain}
\address[CarnegieMellon]{Carnegie Mellon University, Pittsburgh, PA  15213, United States}
\address[Chicago]{Enrico Fermi Institute, University of Chicago, Chicago, Illinois 60637, United States}
\address[Comenius]{Comenius University, 842 48 Bratislava, Slovakia; Institute of Experimental Physics, 040 01 Kosice, Slovakia}
\address[Dubna]{Joint Institute for Nuclear Research, RU-141980 Dubna, Russia}
\address[Duke]{Duke University, Durham, North Carolina  27708, United States}
\address[FNAL]{Fermi National Accelerator Laboratory, Batavia, Illinois 60510, United States}
\address[Florida]{University of Florida, Gainesville, Florida  32611, United States}
\address[Frascati]{Laboratori Nazionali di Frascati, Istituto Nazionale di Fisica Nucleare, I-00044 Frascati, Italy}
\address[Geneva]{University of Geneva, CH-1211 Geneva 4, Switzerland}
\address[Glasgow]{Glasgow University, Glasgow G12 8QQ, United Kingdom}
\address[Harvard]{Harvard University, Cambridge, Massachusetts 02138, United States}
\address[Helsinki]{Division of High Energy Physics, Department of Physics, University of Helsinki and Helsinki Institute of Physics, FIN-00014, Helsinki, Finland}
\address[Illinois]{University of Illinois, Urbana, Illinois 61801, United States}
\address[JohnsHopkins]{The Johns Hopkins University, Baltimore, Maryland 21218, United States}
\address[Karlsruhe]{Institut f\"{u}r Experimentelle Kernphysik, Karlsruhe Institute of Technology, D-76131 Karlsruhe, Germany}
\address[Korea]{Center for High Energy Physics: Kyungpook National University, Daegu 702-701, Korea; Seoul National University, Seoul 151-742, Korea; Sungkyunkwan University, Suwon 440-746, Korea; Korea Institute of Science and Technology Information, Daejeon 305-806, Korea; Chonnam National University, Gwangju 500-757, Korea; Chonbuk National University, Jeonju 561-756, Korea}
\address[LBNL]{Ernest Orlando Lawrence Berkeley National Laboratory, Berkeley, California 94720, United States}
\address[Liverpool]{University of Liverpool, Liverpool L69 7ZE, United Kingdom}
\address[UCLondon]{University College London, London WC1E 6BT, United Kingdom}
\address[Madrid]{Centro de Investigaciones Energeticas Medioambientales y Tecnologicas, E-28040 Madrid, Spain}
\address[MIT]{Massachusetts Institute of Technology, Cambridge, Massachusetts  02139, United States}
\address[Canada]{Institute of Particle Physics: McGill University, Montr\'{e}al, Qu\'{e}bec, H3A~2T8; Simon Fraser University, Burnaby, British Columbia, V5A~1S6; University of Toronto, Toronto, Ontario, M5S~1A7; and TRIUMF, Vancouver, British Columbia, V6T~2A3, Canada}
\address[Michigan]{University of Michigan, Ann Arbor, Michigan 48109, United States}
\address[MichiganState]{Michigan State University, East Lansing, Michigan  48824, United States}
\address[ITEP]{Institution for Theoretical and Experimental Physics, ITEP, Moscow 117259, Russia}
\address[NewMexico]{University of New Mexico, Albuquerque, New Mexico 87131, United States}
\address[Northwestern]{Northwestern University, Evanston, Illinois  60208, United States}
\address[OhioState]{The Ohio State University, Columbus, Ohio  43210, United States}
\address[Okayama]{Okayama University, Okayama 700-8530, Japan}
\address[Osaka]{Osaka City University, Osaka 588, Japan}
\address[Oxford]{University of Oxford, Oxford OX1 3RH, United Kingdom}
\address[Padova]{Istituto Nazionale di Fisica Nucleare, Sezione di Padova-Trento, University of Padova, I-35131 Padova, Italy}
\address[LPNHE]{LPNHE, Universite Pierre et Marie Curie/IN2P3-CNRS, UMR7585, Paris, F-75252 France}
\address[Pennsylvania]{University of Pennsylvania, Philadelphia, Pennsylvania 19104, United States}
\address[Pisa]{Istituto Nazionale di Fisica Nucleare Pisa, University of Pisa, University of Siena and Scuola Normale Superiore, I-56127 Pisa, Italy}
\address[Pittsburgh]{University of Pittsburgh, Pittsburgh, Pennsylvania 15260, United States}
\address[Purdue]{Purdue University, West Lafayette, Indiana 47907, United States}
\address[Rochester]{University of Rochester, Rochester, New York 14627, United States}
\address[Rockefeller]{The Rockefeller University, New York, New York 10021, United States}
\address[Roma]{Istituto Nazionale di Fisica Nucleare, Sezione di Roma 1, Sapienza Universit\`{a} di Roma, I-00185 Roma, Italy}
\address[Rutgers]{Rutgers University, Piscataway, New Jersey 08855, United States}
\address[TexasAM]{Texas A\&M University, College Station, Texas 77843, United States}
\address[Trieste]{Istituto Nazionale di Fisica Nucleare Trieste/Udine, I-34100 Trieste, University of Trieste/Udine, I-33100 Udine, Italy}
\address[Tsukuba]{University of Tsukuba, Tsukuba, Ibaraki 305, Japan}
\address[Tufts]{Tufts University, Medford, Massachusetts 02155, United States}
\address[Waseda]{Waseda University, Tokyo 169, Japan}
\address[WayneState]{Wayne State University, Detroit, Michigan  48201, United States}
\address[Wisconsin]{University of Wisconsin, Madison, Wisconsin 53706, United States}
\address[Yale]{Yale University, New Haven, Connecticut 06520, United States}
\author[Helsinki]{T.~Aaltonen}
\author[Chicago]{J.~Adelman}
\author[Cantabria]{B.~\'{A}lvarez~Gonz\'{a}lez\fnref{fn1}}
\author[Padova]{S.~Amerio}
\author[Michigan]{D.~Amidei}
\author[Northwestern]{A.~Anastassov}
\author[Frascati]{A.~Annovi}
\author[Comenius]{J.~Antos}
\author[FNAL]{G.~Apollinari}
\author[Purdue]{A.~Apresyan}
\renewcommand{\thefootnote}{\fnsymbol{footnote}}
% ...

\begin{abstract}
We present the result of a search for a massive color-octet vector particle,
(e.g. a massive gluon) decaying to a pair of top quarks in proton-antiproton
collisions with a center-of-mass energy of 1.96~TeV. 
This search is based on 1.9~fb$^{-1}$ of data collected using the CDF detector
during Run II of the Tevatron at Fermilab.
We study $t\bar{t}$ events in the lepton+jets channel with at least one
$b$-tagged jet.
A massive gluon is characterized by its mass, decay width, and the strength of
its coupling to quarks.
These parameters are determined according to the observed invariant mass distribution
of top quark pairs.
We set limits on the massive gluon coupling strength for masses between 400 and 800
GeV$/c^2$ and width-to-mass ratios between 0.05 and 0.50.
The coupling strength of the hypothetical massive gluon to quarks is consistent
with zero within the explored parameter space.
\end{abstract}



\end{frontmatter}

\section{Introduction}
\label{intro}
The top quark is the heaviest known elementary particle, with a mass very
close to the electroweak symmetry-breaking scale. 
As such, the top could be sensitive to physics beyond the standard model (SM)
\cite{bib:Hill1,bib:Hill2}. 
New particles decaying to $t\bar{t}$ pairs can be scalar or vector,
color-singlet or color-octet; 
a scalar resonance is predicted in two-Higgs-doublets models
\cite{bib:2HDM1,bib:2HDM2};
vector particles appear as massive $Z$-like bosons in extended gauge theories
\cite{bib:Zlike}, or as Kaluza-Klein states of the gluon and $Z$ boson
\cite{bib:KKgluon,bib:KKZ}, or as colorons \cite{bib:Hill1}. 
Searches for a color-singlet particle decaying to a $t\bar{t}$ pair have been
performed by CDF and D0 collaborations in Run I \cite{bib:CDFRunI,bib:D0RunI}
and Run II \cite{bib:CDFMtt1,bib:CDFMtt2,bib:D0Mtt}.

\begin{figure}[H]
 \begin{center}
  \includegraphics[width=2in]{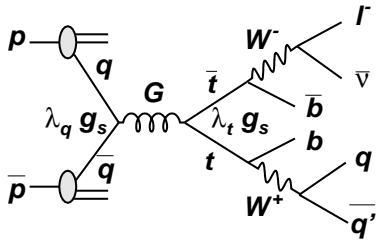} 
  \caption{A Feynman diagram showing $t\bar{t}$ production via a massive gluon
    in the ``lepton+jets'' decay channel.}
  \label{fig:feynmann}
 \end{center}
\end{figure}

In this Letter we describe a search for a massive color-octet vector particle
{$G$}, which we call generically a ``massive gluon''.
We assume $G$ has a much stronger coupling to the top quark than 
to the lighter quarks, $q=(u,d,c,s,b)$ \cite{bib:Hill1}. 
We also assume the massive-massless gluon coupling is negligible. 
A Feynman diagram for $t\bar{t}$ production via massive-gluon 
is shown in Fig.~\ref{fig:feynmann}. 
The coupling strength of massive gluons to light and top quarks are 
$\lambda_{q} g_{s}$ and $\lambda_{t}g_{s}$, respectively, where 
$g_{s}$ is the coupling constant of the SM strong interaction.

In $t\bar{t}$ production, there are three observable parameters: the product of 
massive gluon coupling strength  $\lambda\equiv \lambda_{q}\lambda_{t}$, mass $M$, 
and  width $\Gamma$.
In this analysis we consider only the possibility of novel strong-sector
production; we assume that the weak decay of the top quark follows the SM. 
Since the color and the current structures of $G$ and SM gluon ($g$) are identical, 
interference between processes through massive gluon and massless gluon will produce a
$t\bar{t}$ invariant mass distribution with an enhanced signal that has a
characteristic form  \cite{bib:Hill1} as shown later in Fig.~\ref{fig:ttinvmass2}.
If the coupling of $G$ to quarks is assumed to be parity-conserving, the
production matrix element can be written as

\begin{eqnarray}
|\mathcal{M}_{\rm prod.}|^2 &=& \frac{2}{9} g_s^4 \hat{s}^2 (2 - \beta^2 + \beta^2 \cos^2 \theta) (  \Pi_g  + \lambda \Pi_{\rm int.} + \lambda^2 \Pi_G),
\end{eqnarray}
where $\hat{s}$, $\beta$, and $\theta$ are the $t\bar{t}$ invariant mass squared, 
the velocity of the top quark, and the angle between the top quark and the incident quark in the $t\bar{t}$ center of mass system, respectively.
The propagator factors of gluons, massive gluons, and interference are

\begin{eqnarray}
\Pi_g = \frac{1}{\hat{s}^2}, \
\Pi_G = \frac{1}{(\hat{s} - M^2)^2 + M^2\Gamma^2}, \
\Pi_{\rm int.} = \frac{2}{\hat{s}}\frac{\hat{s}-M^2}{(\hat{s} - M^2)^2 + M^2\Gamma^2}.
\end{eqnarray}
We search for $t\bar{t}$ pairs produced by massive and massless gluons, 
where interference between these diagrams (denoted by $g+G$) is considered, by
examining the invariant mass spectrum of observed $t\bar{t}$ candidate events.

\section{Selection of candidate events}
\label{selection}
The search is based on data collected with the CDF II detector between March
2002 and May 2007 at the Fermilab Tevatron $p\bar{p}$ collider,
corresponding to an integrated luminosity of about 1.9 fb$^{-1}$. 
The CDF II detector is a general purpose detector which is azimuthally and
forward-backward symmetric. 
The detector consists of a charged particle tracking system composed of silicon
microstrip detectors and a gas drift chamber inside a 1.4 T magnetic field,
surrounded by electromagnetic and hadronic calorimeters and enclosed by muon
detectors.
The details of the detector are described elsewhere
\cite{bib:CDFII_SecVtxBtag_MethodII}. 

The cross section for standard model $t\bar{t}$ production in $p\bar{p}$
collisions at $\sqrt{s}=1.96$~TeV is dominated by $q\bar{q}$ annihilation ($\sim
85 \%$). The remaining $\sim 15\%$ is attributed to gluon-gluon
fusion~\cite{bib:SMttXsec}. 
Standard model top quarks decay almost exclusively to $Wb$.
The search presented here focuses on the $t\bar{t}$ event topology wherein one
$W$ boson decays hadronically while the other decays to an electron or muon and
its corresponding neutrino.
Both $b$ quarks and the two decay quarks of the second $W$ boson appear as jets in
the detector. 
Accordingly, $t\bar{t}$ candidates in this ``lepton + jets'' channel are
characterized by a single lepton, missing transverse energy
$\met$~\cite{bib:CDFcoodinate_met}, due to the undetected neutrino, and four
jets.  

We use lepton triggers that require an electron or muon with $p_T > 18$~GeV/$c$. Events included in our analysis must first contain an isolated electron (muon)
with $E_T>20$~GeV ($p_T>20$~GeV$/c$) in the central detector region
with $|\eta|<1.0$. 
Electron and muon identification methods are described in
Ref.~\cite{bib:LeptonID}. 
We remove events which have multiple leptons, cosmic ray muons, electrons from
photon-conversion, or tracks, such that its momenta added to the primary lepton momenta gives an invariant mass equal to the $Z$ mass.  
The position of the primary vertex (along the beam) is required to be within 60
cm of the center of the nominal beam intersection and consistent with the reconstructed $z$
position of the high-$p_T$ lepton. 
Events must also feature at least 20~GeV of $\met$ attributable to the presence
of a high-$p_T$ neutrino, as well as exactly four jets with $|\eta|<2.0$ and
$E_T>20$~GeV (jet $E_T$ corrections are described in Ref. \cite{bib:JECR}).
Jets are clustered with a cone-based algorithm, with a cone size 
$\Delta R \equiv \sqrt{\Delta\phi^2 + \Delta\eta^2}=0.4$.
We reduce non-$t\bar{t}$ backgrounds by requiring at least one jet
identified by the displaced secondary vertex ``$b$-tagging''
algorithm~\cite{bib:CDFII_SecVtxBtag_MethodII} as being consistent with the
decay of a long-lived $b$ hadron. 

\section{Background}
\label{background}
The background to the $t\bar{t}$ signal is mostly from $W$+jets events with a
falsely-reconstructed secondary vertex (Mistags) or from $W$+jets events where
one or more jets are due to heavy-flavor quarks.
Smaller contributions are from QCD multi-jet production, 
in which either the $W$ signature is faked when jets contain
 semi-leptonic $b$-hadron decays or when jets are mis-reconstructed
 and appear as electrons and missing $E_T$, single top quark production, 
diboson ($WW$, $WZ$, $ZZ$) production, 
and $Z$ bosons produced in association with multiple jets.
The methods used to estimate the backgrounds are detailed in
Ref. \cite{bib:CDFII_SecVtxBtag_MethodII}. 
The $gg\rightarrow t\bar{t}$ process is taken as a background for this search,
which is estimated by assuming $0.16 \pm 0.05$ gluon fusion fraction from
Ref. \cite{bib:SMttXsec}. 
The estimated backgrounds in the sample are summarized in Table~\ref{tab:BackgroundComposition}.
The diboson and $gg\rightarrow t\bar{t}$
background processes are modeled with {\sc pythia} \cite{bib:Pythia} , $W$+jets and $Z$+jets
processes with {\sc alpgen} \cite{bib:Alpgen}, and QCD with data.

\begin{table}
  \begin{center}
    \caption{Background composition to the $q\overline{q} \to t\overline{t}$
      process and the expected numbers of events. Systematic uncertainties
      coming from the background estimation method are listed. Electroweak
      includes single top quark, diboson production, and $Z$ bosons + jets
      productions. A luminosity of 1.9 fb$^{-1}$ is assumed.} 
\vspace*{8mm}
    \begin{tabular}{lc} \hline\hline
	Source & Expected number of events \\
	\hline
	Electroweak                     & $9.9 \pm 0.5$   \\
	$W$ + bottom                    & $16.5 \pm 6.7$  \\
	$W$ + charm                     & $12.9 \pm 5.2$  \\
	Mistags                         & $16.7 \pm 3.6$  \\
	non-$W$                        & $13.6 \pm 11.7$ \\
	SM $gg\rightarrow t\bar{t}$        & $48.2 \pm 15.6$ \\
	\hline
	Total Background ($n_{b}^{\textrm{exp}}\pm\sigma_{b}^{\textrm{exp}}$) & $117.8 \pm 19.8$ \\
	SM $q\bar{q}\rightarrow t\bar{t}$ ($\sigma=5.6$ pb) & $211.7\pm 29.3$ \\
	\hline
	Data & 371 \\
	\hline\hline
    \end{tabular}
    \label{tab:BackgroundComposition}
  \end{center}
\end{table}

\section{$t\bar{t}$ invariant mass reconstruction}
\label{reconstruction}
We search for a massive gluon by examining the $t\bar{t}$ invariant mass
$(\sqrt{\hat{s}}_{t\bar{t}})$ spectrum of the selected events. 
Our analysis procedure consists of two steps.  \\
(A) We first reconstruct the parton level momentum set of $t\bar{t}=(bl^{+}\nu)(\bar{b}l^{-}\bar{\nu})$, event-by-event, with the Dynamical Likelihood Method (DLM) \cite{bib:DLM,bib:DLMMtop}. 
The likelihood in DLM is defined by the differential cross section per unit phase space volume, multiplied by a posterior transfer function (TF),
which is a probability density function from the observed to the parton
kinematic quantities. For a given event, different parton kinematics set are inferred with TF.
We assume in this paper the TF is independent of the production matrix for $t\bar{t}$, hence in the reconstruction, we remove the $t\bar{t}$ production matrix  from the likelihood. 
For each event the reconstructed $\sqrt{\hat{s}}_{t\bar{t}}$ is averaged over inferred parton paths, and denoted by  $\InvMtt_r$. We denote the true $\sqrt{\hat{s}}_{t\bar{t}}$ at the parton level by $\InvMtt_p$. \\
(B) We interpret $\InvMtt_r$ as an observed quantity and apply a standard
unbinned likelihood technique to reproduce the $\InvMtt_r$ distribution using Monte Carlo (MC) events, which consist of signal and relevant background processes.

\begin{figure}[H]
 \begin{center}
  \includegraphics[width=2.5in]{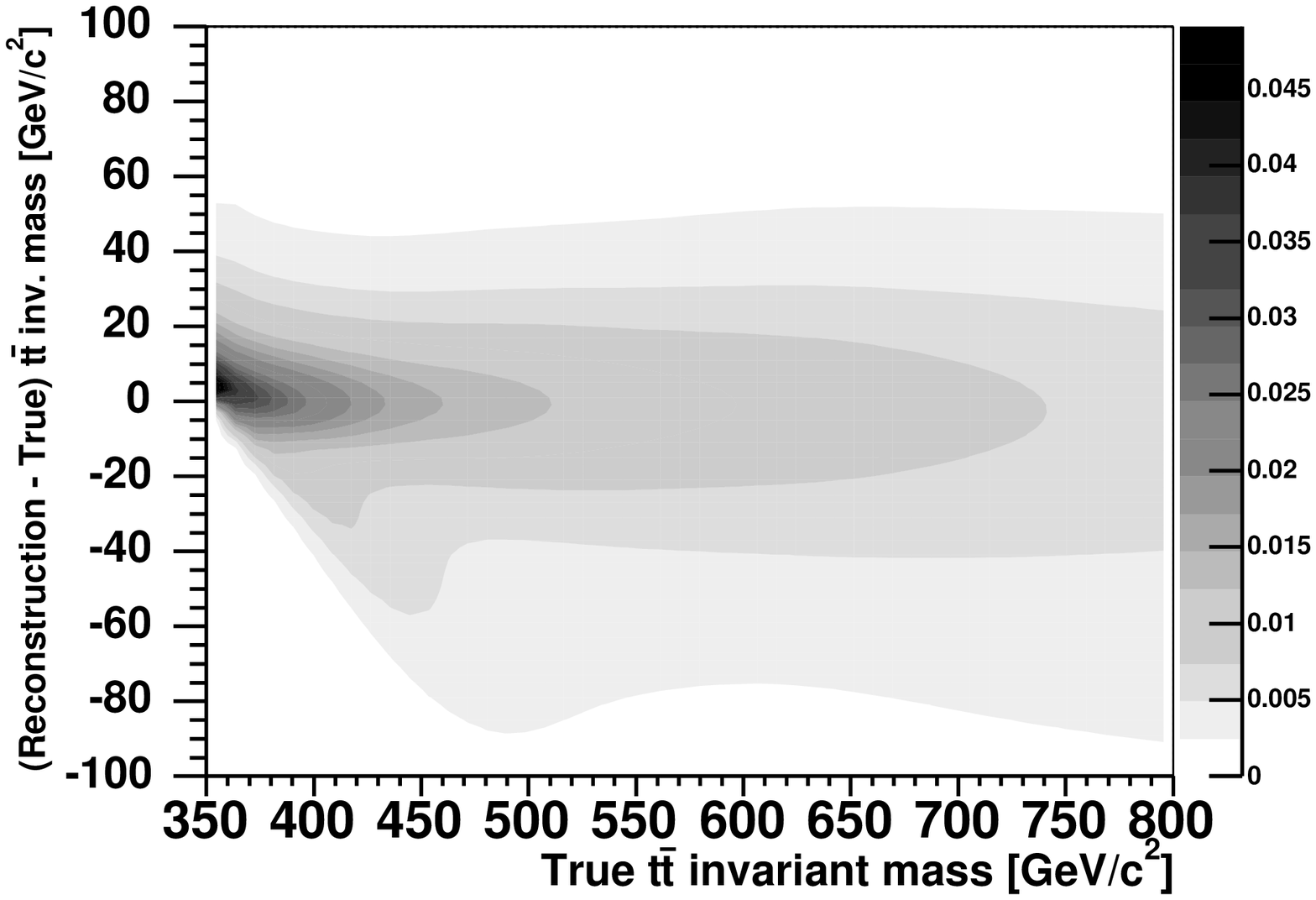} 
  \caption{Resolution function 
$f(y;x)$, where $x = \InvMtt_p$ and $y=\InvMtt_r - \InvMtt_p$. 
}
  \label{fig:ttinvmass1}
 \end{center}
\end{figure}

The $\InvMtt_r$ distribution for the signal $g+G$ process is expressed relative to $[d\sigma/d{\InvMtt_p}]^*_{SM:\qqtt}$, the SM cross section after the event selection cuts (expressed by $^*$) by

\begin{eqnarray}
p_s [\InvMtt_r; \boldsymbol{\alpha}] \equiv N(\boldsymbol{\alpha}) \int \left[ \frac{d\sigma}{d\InvMtt_p} \right]_{SM:\qqtt}^* R(\InvMtt_p; \boldsymbol{\alpha}) f(\InvMtt_r - \InvMtt_p; \InvMtt_p)d\InvMtt_p ,
\label{eq:SignalInvMttPDF}
\end{eqnarray}
where $\boldsymbol{\alpha}\equiv(\lambda, M, \Gamma)$ is a set of massive gluon
parameters, and $N(\bm{\alpha})$ is the normalization factor.
The ratio of $g+G$  to SM $t\overline{t}$ production cross sections is

\begin{eqnarray}
R(\InvMtt_p; \boldsymbol{\alpha}) &=& 
 1 + 2\lambda \frac{ \InvMttSQp(\InvMttSQp-M^2) }{(\InvMttSQp-M^2)^2 +
   M^2\Gamma^2} + \lambda^2 \frac{\InvMttSQp^2}{(\InvMttSQp-M^2)^2 + M^2
   \Gamma^2}. \label{eq:prpRatio}
\end{eqnarray}
In this ratio, PDF effects, top propagators, decay matrix elements and the final state
densities for $g+G$ and SM processes cancel out, making it possible to
generate (or derive) $g+G$ events with specified $\bm{\alpha}$ from standard
model $t\bar{t}$ MC events.
The resolution function $f$ translates $\InvMtt_p$ to $\InvMtt_r$.
It is obtained from the scatter plot of $\InvMtt_r - \InvMtt_p$ vs. $\InvMtt_p$
using the {\sc pythia} SM $q\bar{q}\rightarrow t\bar{t}$ MC sample, as shown in
Fig.~\ref{fig:ttinvmass1}.
The peak and half maximum points of $\InvMtt_r - \InvMtt_p$ are typically
$-0.2^{+15.7}_{-13.7}$, $-0.9^{+31.0}_{-40.7}$, and $-3.5^{+39.6}_{-68.7}$
GeV/$c^2$ at $\sqrt{s_{p}}=400$, $600$, and $800$~GeV/$c^2$, respectively.

\begin{figure}[H]
 \begin{center}
  \includegraphics[width=2in]{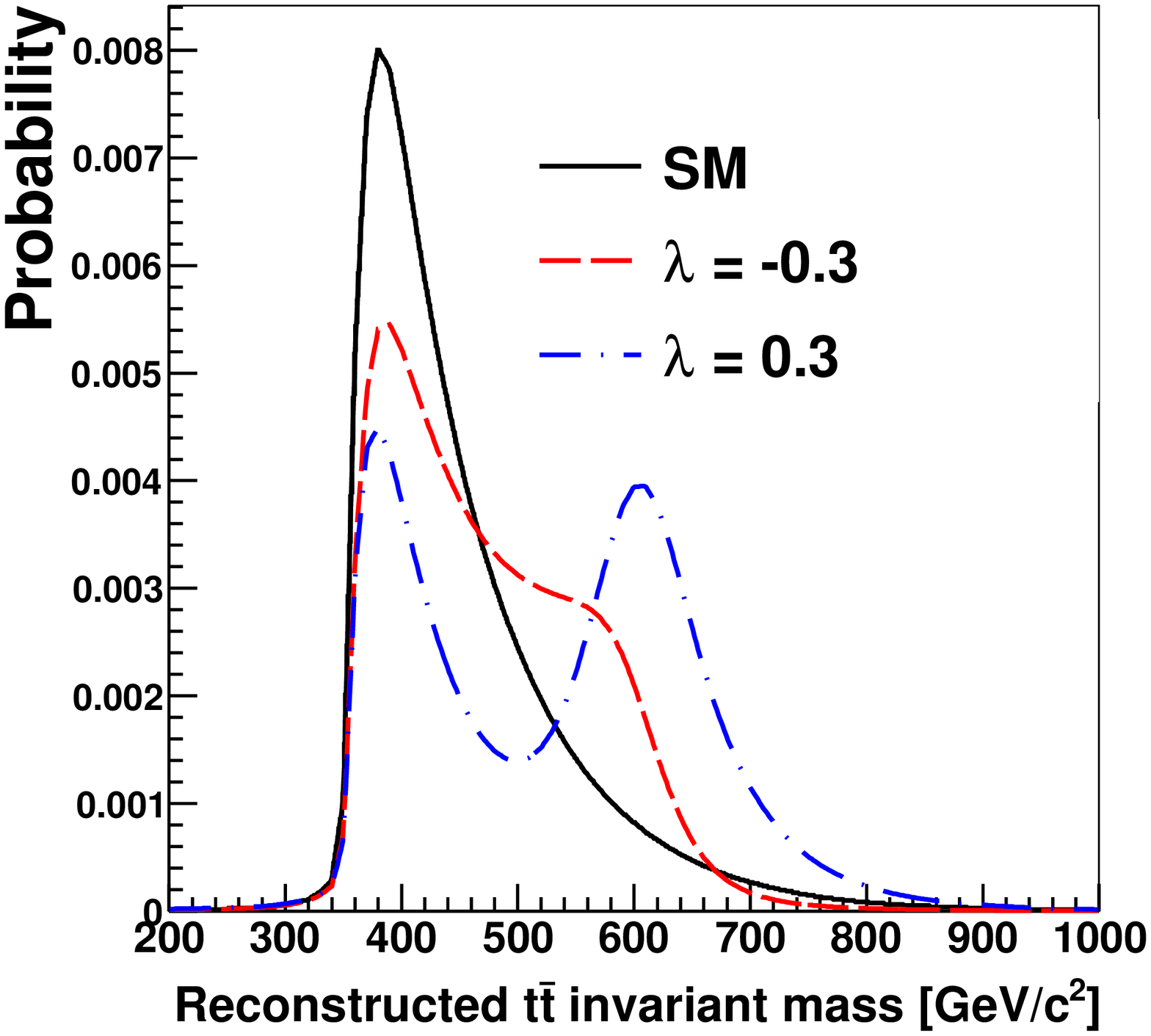}
  \caption{Signal probability density function $p_s [\InvMtt_r; \boldsymbol{\alpha}]$ with $M=600$~GeV/$c^2$, $\Gamma/M=0.10$, and $\lambda=\pm0.3$.}
  \label{fig:ttinvmass2}
 \end{center}
\end{figure}

Figure~\ref{fig:ttinvmass2} shows an example of $p_s [\InvMtt_r;
\boldsymbol{\alpha}]$.  
The first(second) peak is due to $g(G)$. 
At the parton level, the $G$ peak(dip) is above(below) $M$ for
positive(negative) $\lambda$, while at the reconstructed $\InvMtt_r$
distribution the peak(dip) is shifted due to the resolution function.
The {\sc pythia} MC with $M_t=175$~GeV/$c^2$ is used to estimate
$[d\sigma/d{\InvMtt_p}]^*_{SM:\qqtt}$.

\section{Extraction of coupling strength}
\label{extraction}
If $M$ and $\Gamma/M$ are given, the  $\InvMtt_r$ distribution of the data is a
function of $\lambda$ and the numbers of the signal and background events,
$n_{s}$ and $n_{b}$. 
By applying a three parameter unbinned likelihood method for $(\lambda, n_{s},
n_{b})$, the likelihood function $L$ for the distribution is 

{\small
\begin{eqnarray}
L \left(\lambda, n_s, n_b | M, \frac{\Gamma}{M} \right) \equiv G(n_b; n_b^{\textrm{exp}}, \sigma_b^{\textrm{exp}} ) P(N;n) \prod_{i=1}^{N} \frac{n_s p_s(\InvMtt_r(i);\boldsymbol{\alpha}) +n_b p_b(\InvMtt_r(i)) }{n}.\label{eq:LambdaLikelihood}
\end{eqnarray}
}
Function $G(n_b; n_b^{\textrm{exp}}, \sigma_b^{\textrm{exp}} )$ constrains the
total background normalization with Gaussian probability around the central
value $n_b^{\textrm{exp}}$ and its uncertainty $\sigma_b^{\textrm{exp}}$ as
given in Table~\ref{tab:BackgroundComposition}.
$P(N;n)$  is the Poisson probability for observing $N$ events where $n=n_s +
n_b$ are expected.  
$p_s(\InvMtt_r(i);\boldsymbol{\alpha})$ and $p_b(\InvMtt_r(i))$ are
probability densities that the $i$-th event is due to signal or background,
respectively. 
We calculate the likelihood as a function of $\lambda$ for a number of values of
$(M,\Gamma/M)$,  
where the value of $M$ ranges from 400 to 800~GeV/c$^{2}$ at 50~GeV/c$^{2}$
intervals. $\Gamma/M$ is considered at values of  0.05, 0.1, 0.2, 0.3, 0.4 and
0.5.  

\begin{figure}[H]
 \begin{center}
  \includegraphics[width=2.7in]{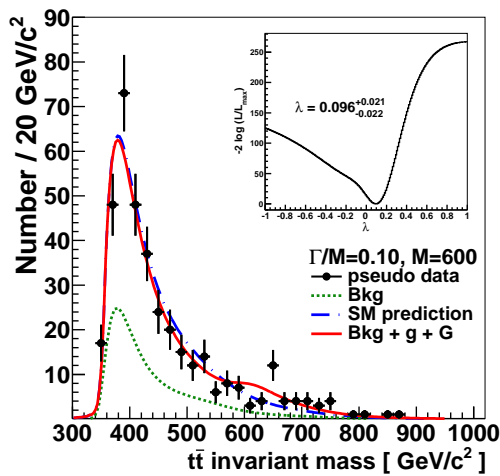}
  \caption{An example of a likelihood fit to $\sqrt{\hat{s}_r}$ spectrum for 1.9
    fb$^{-1}$ MC events (points), which includes the massive gluon with a
    $\Gamma/M=0.1$, $M=600$~GeV/$c^2$, and $\lambda=0.1$. The solid
    line is the best fit to a superposition of $g+G$ signal and the expected
    background (dash line) and SM prediction (dot-dash line). The likelihood fit
    is performed assuming $\Gamma/M=0.1$ and $M=600$~GeV/$c^2$. The inset shows
    the $-2 \ln L$ as a function of $\lambda$.}
  \label{fig:PE}
 \end{center}
\end{figure}

The analysis method is tested with pseudo-experiments~(PE's), where the
background events are generated according to
Table~\ref{tab:BackgroundComposition}, and the total number of events is
normalized to the 371 observed number of candidate events.
An example of the analysis of a single pseudo-experiment is shown in
Fig.~\ref{fig:PE}.
Points represent the observed $\sqrt{\hat{s}}$ distribution for a simulated signal
sample with $\Gamma/M=0.1$, $M=600$~GeV/$c^2$ and $\lambda=0.1$. 
$\lambda$ is estimated by maximizing the likelihood $L$, given the
observed $\InvMtt_r$ spectrum.
The 95\% C.L. lower/upper two-sided limit on $\lambda$ at the given mass and
width is found by taking $-2 \ln(L/L_{max}) =3.84$. 
We measure our expected sensitivity using a large number of null signal PE's.

We have studied contributions to the total uncertainty arising from systematic
effects.
Variation of the $t\bar{t}$ invariant mass distribution affects the
estimation of the coupling strength, $\lambda$.
The jet energy scale and top-quark mass are varied simultaneously (within their
uncertainties) to properly account for their correlation. 
The uncertainties of the parton distribution functions (PDF) are estimated by
comparing PE's where different PDF sets (CTEQ5L\cite{bib:CTEQ} vs MRST72\cite{bib:MRS}) are
used; additionally, $\Lambda_{QCD}$ and the CTEQ6M\cite{bib:CTEQ} PDF eigenvectors are
varied.
The systematic effect due to choice of MC generator is estimated by comparing
PE's  using events generated by {\sc pythia} and {\sc herwig}~\cite{bib:Herwig}.
The scale of next-to-leading-order systematic effects is estimated by using
events generated with {\sc mc@nlo}\cite{bib:mcnlo}.  
Systematic uncertainties in modeling initial- and final-state gluon radiation
are estimated using {\sc pythia}, where the range of QCD parameters are varied
in accordance with CDF studies of Drell-Yan data~\cite{bib:isrfsr}. 
The uncertainty in the MC modeling of the multiple interaction and the
$b$-tagging efficiency as a function of jet $p_T$ are evaluated.  
All systematic uncertainties are evaluated at the full range of coupling
strengths, gluon masses and decay widths considered, and are incorporated in the
likelihood function.

\section{Results}
\label{results}
\begin{figure}[H]
 \begin{center}
  \includegraphics[width=2.5in]{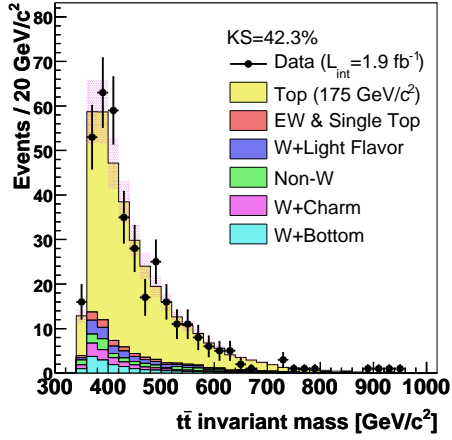}
  \caption{The $t\bar{t}$ invariant mass distribution. The points represent
    the observed distribution; the histogram represents the SM prediction
    normalized to the observed data. 
    The pink bands represent the uncertainty on the background estimations.
  } 
  \label{fig:ttinvmass}
 \end{center}
\end{figure}

\begin{figure}[H]
 \begin{center}
  \includegraphics[width=2.2in] {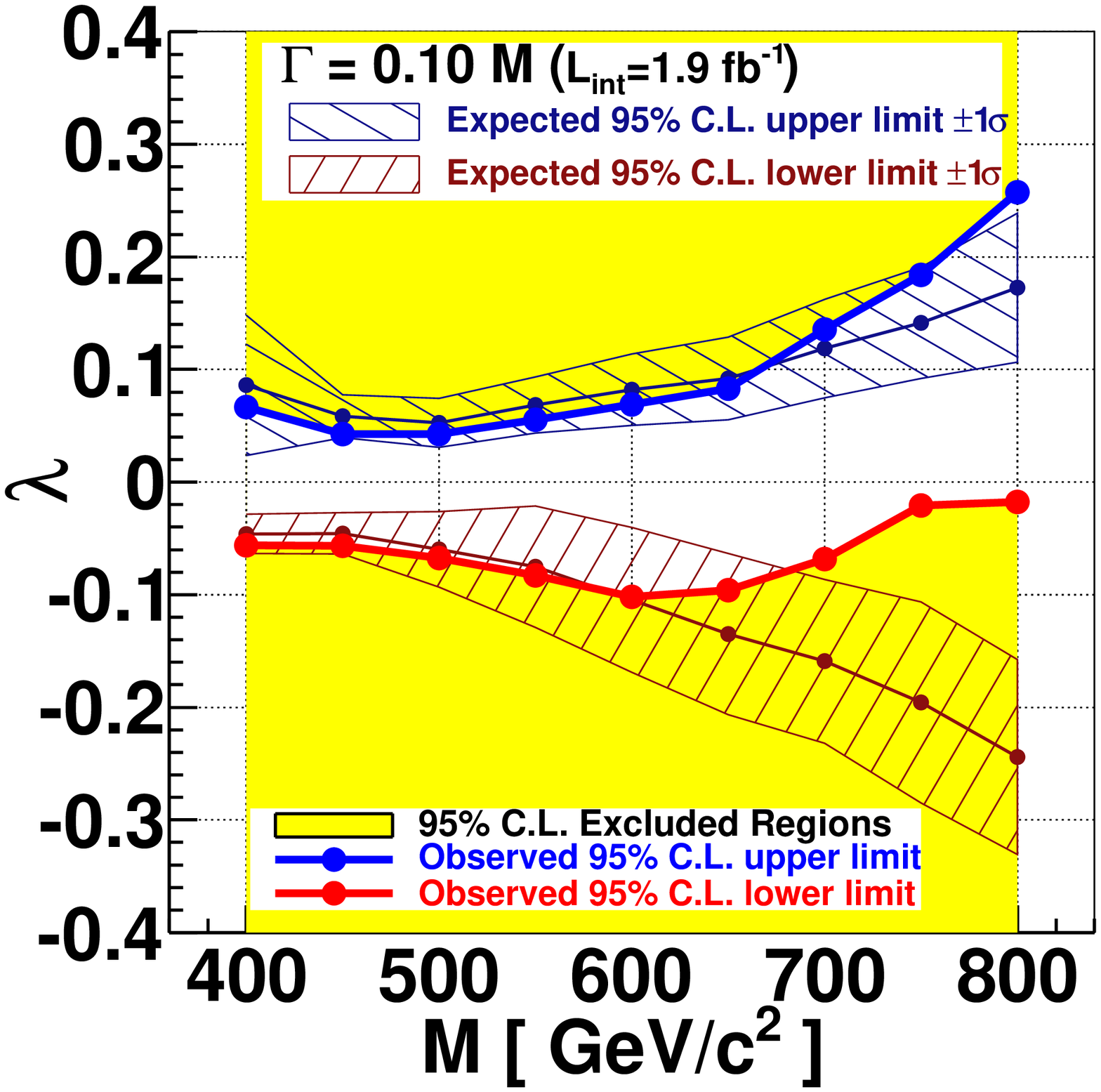}
  \includegraphics[width=2.2in] {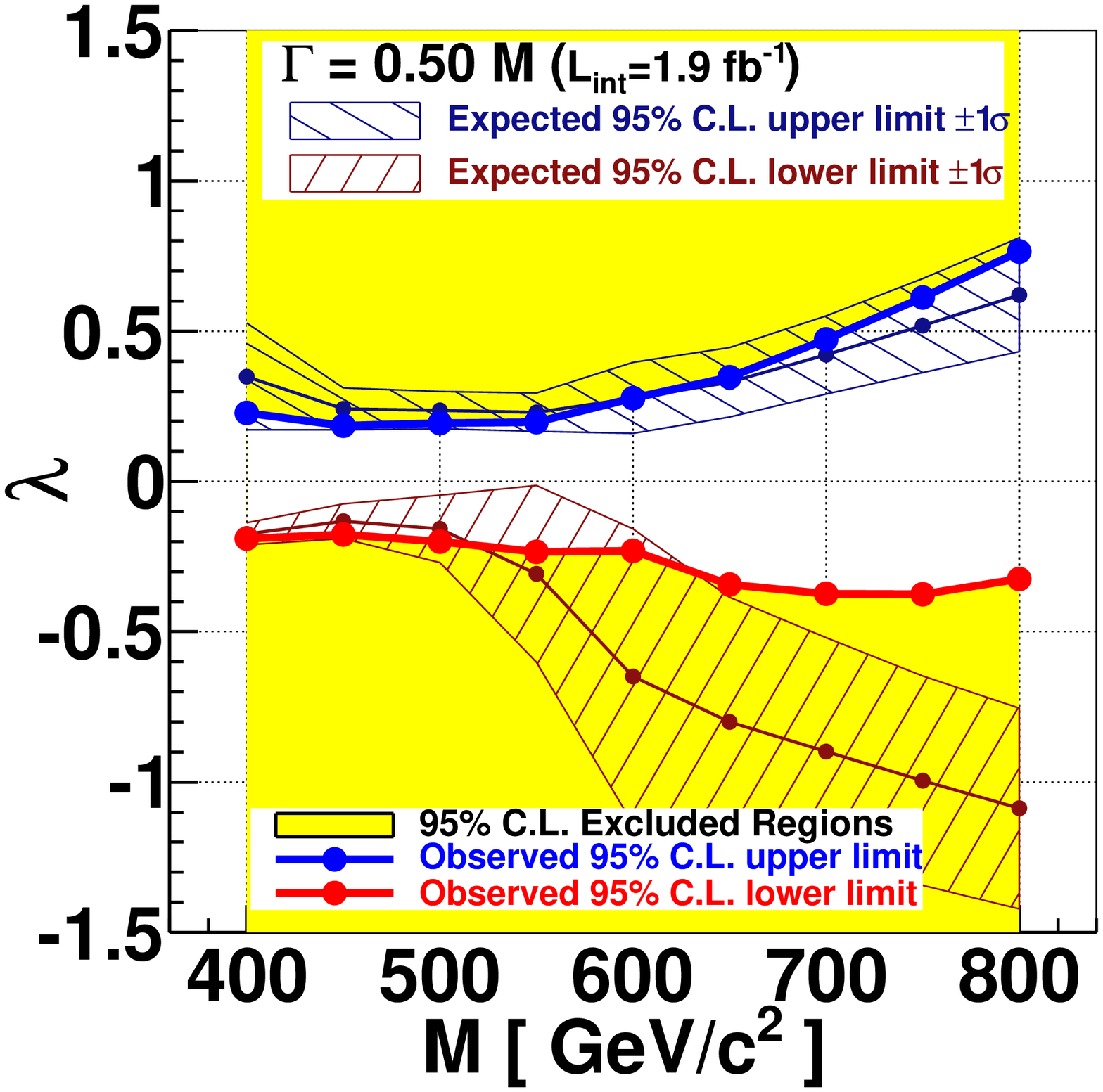}
  \caption{Excluded regions of $\lambda$ in the $\lambda$-$M$ plane (yellow), for $\Gamma/M = 0.1$ (left) and $\Gamma/M = 0.5$ (right). The expected limit is defined as median 95\% value and $\pm 1 \sigma$ is defined with regard to the median for the set of null signal PE's.}
  \label{fig:limit}
 \end{center}
\end{figure}

The $t\bar{t}$ invariant mass distribution observed in data is shown in
Fig.~\ref{fig:ttinvmass}.  
The best fit of $\lambda$ is consistent with the SM prediction, including
a fluctuation of $\sim1.7\sigma$ within the explored parameter range.
We search for massive gluon with mass in the range [400,800]~GeV/c$^{2}$, 
though events with gluon masses beyond this region are included in the analysis. 
The 95\% C.L. limits on the coupling strength $\lambda$ at $\Gamma/M=0.1$ and
$\Gamma/M=0.5$ are shown in Fig.~\ref{fig:limit}.  
A less stringent limit above 650~GeV/$c^2$ is due to 4 events on the high
mass tail. The SM predicts 1.4 events above 850~GeV/$c^2$, 4 events observed
correspond to a Poisson probability of 5.4~\%.
The limits at 95$\%$ C.L. for several values of $M$ and $\Gamma/M$  are listed
in Table~\ref{tab:LowerUpperLimits}.
The limits become weaker with higher $M$ and wider $\Gamma/M$.\\
In conclusion, we peform an exploratory search for a color-octet vector particle
in general with minimal model dependence.
No significant indication of the existence of massive gluon  with $|\lambda|>0.5$ is
observed in our search region of 400~GeV/c$^{2}$ $< M <$ 800~GeV/c$^{2}$ and 0.05
$< \Gamma/M <$ 0.5. 

{\tiny
\begin{table}[H]
 \begin{center}
  \caption{Expected and observed 95\% C.L. lower/upper limits on $\lambda$. Expected limits are in parentheses.}
  \begin{tabular}{ccccccc}
   \hline
   \hline
   &   $\Gamma/M=0.05$ &   $\Gamma/M=0.10$ &   $\Gamma/M=0.20$ &   $\Gamma/M=0.30$ &   $\Gamma/M=0.40$ &  $\Gamma/M=0.50$ \\
   \hline
   M=400 & (-0.036 / 0.042) & (-0.046 / 0.086) & (-0.074 / 0.156) & ( -0.11 / 0.26) & (-0.15 / 0.30) & (-0.17 / 0.35) \\
         & -0.043 / 0.040 & -0.056 / 0.067 & -0.089 / 0.11 & -0.12 / 0.18 & -0.16 / 0.20 & -0.19 / 0.23 \\
   M=450 & (-0.038 / 0.040) & (-0.046 / 0.058) & (-0.065 / 0.087) & (-0.087 / 0.13) & (-0.12 / 0.19) & (-0.13 / 0.24) \\
         & -0.045 / 0.027 & -0.057 / 0.042 & -0.086 / 0.06 & -0.12 / 0.09 & -0.15 / 0.14 & -0.18 / 0.19 \\
   M=500 & (-0.051 / 0.038) & (-0.060 / 0.053) & (-0.083 / 0.087) & ( -0.12 / 0.13) & (-0.13 / 0.18) & (-0.16 / 0.24) \\
         & -0.059 / 0.034 & -0.067 / 0.043 & -0.10  / 0.06 & -0.14 / 0.10 & -0.17 / 0.14 & -0.20 / 0.19 \\
   M=550 & (-0.058 / 0.049) & (-0.075 / 0.069) & ( -0.13 / 0.10 ) & ( -0.15 / 0.15) & (-0.20 / 0.20) & (-0.31 / 0.23) \\
         & -0.064 / 0.039 & -0.083 / 0.055 & -0.12  / 0.08 & -0.16 / 0.13 & -0.19 / 0.18 & -0.23 / 0.20 \\
   M=600 & (-0.074 / 0.058) & ( -0.10 / 0.082) & ( -0.19 / 0.12 ) & ( -0.21 / 0.18) & (-0.35 / 0.22) & (-0.65 / 0.28) \\
         & -0.073 / 0.048 & -0.10  / 0.069 & -0.16  / 0.10 & -0.15 / 0.16 & -0.22 / 0.20 & -0.23 / 0.28 \\
   M=650 & (-0.098 / 0.077) & ( -0.14 / 0.092) & ( -0.24 / 0.15 ) & ( -0.36 / 0.20) & (-0.58 / 0.27) & (-0.80 / 0.33) \\
         & -0.081 / 0.069 & -0.096 / 0.083 & -0.13  / 0.15 & -0.16 / 0.20 & -0.26 / 0.29 & -0.34 / 0.35 \\
   M=700 & ( -0.11 / 0.082) & ( -0.16 / 0.12 ) & ( -0.27 / 0.17 ) & ( -0.44 / 0.25) & (-0.68 / 0.32) & (-0.90 / 0.42) \\
         & -0.070 / 0.091 & -0.068 /  0.14 & -0.091 / 0.19 & -0.13 / 0.29 & -0.31 / 0.37 & -0.37 / 0.47 \\
   M=750 & ( -0.14 / 0.11 ) & ( -0.20 / 0.14 ) & ( -0.33 / 0.21 ) & ( -0.50 / 0.31) & (-0.71 / 0.39) & (-1.00 / 0.52) \\
         & -0.020 /  0.13 & -0.021 /  0.18 & -0.033 / 0.26 & -0.03 / 0.38 & -0.06 / 0.47 & -0.37 / 0.61 \\
   M=800 & ( -0.16 / 0.12 ) & ( -0.24 / 0.17 ) & ( -0.39 / 0.27 ) & ( -0.59 / 0.37) & (-0.80 / 0.49) & (-1.09 / 0.62) \\
         & -0.011 /  0.18 & -0.017 /  0.26 &  0.017 / 0.37 & -0.04 / 0.50 & -0.01 / 0.63 & -0.33 / 0.76 \\
   \hline
   \hline
  \end{tabular} 
  \label{tab:LowerUpperLimits}
 \end{center}
\end{table}
}

\section*{Acknowledgements}
\label{Acknow}
We thank the Fermilab staff and the technical staffs of the participating institutions for their vital contributions. This work was supported by the U.S. Department of Energy and National Science Foundation; the Italian Istituto Nazionale di Fisica Nucleare; the Ministry of Education, Culture, Sports, Science and Technology of Japan; the Natural Sciences and Engineering Research Council of Canada; the National Science Council of the Republic of China; the Swiss National Science Foundation; the A.P. Sloan Foundation; the Bundesministerium f\"ur Bildung und Forschung, Germany; the World Class University Program, the National Research Foundation of Korea; the Science and Technology Facilities Council and the Royal Society, UK; the Institut National de Physique Nucleaire et Physique des Particules/CNRS; the Russian Foundation for Basic Research; the Ministerio de Ciencia e Innovaci\'{o}n, and Programa Consolider-Ingenio 2010, Spain; the Slovak R\&D Agency; and the Academy of Finland. 

\section*{References}
\label{Refs}


\begin{thebibliography}{99}
\bibitem{bib:CDFcoodinate_met} We use a coordinate system defined about the proton beam direction, which is taken as the $z$ axis; the $x$ axis lies in the horizontal plane. Then $\theta$ is the usual polar angle and $\phi$ is the azimuthal angle. We define the pseudorapidity $\eta$ of a particle's three-momentum as $\eta \equiv -\ln(\tan\frac{\theta}{2})$. The transverse energy and momentum are defined as $E_T = E \sin\theta$ and $p_T = p\sin\theta$ where $E$ is the energy measured by the calorimeter and $p$ is the momentum measured in the tracking system. The missing transverse energy is defined as $\met = |-\sum_i E_T^i \vec{n}_i |$ where $\vec{n}_i$ is a unit vector in the transverse plane that points from the event vertex to the azimuth of the $i^{\rm th}$ calorimeter tower.
\end{thebibliography}
\end{document}